# Dielectric Engineering of Electronic Correlations in a van der Waals Heterostructure

Philipp Steinleitner[1], Philipp Merkl[1], Alexander Graf[1], Philipp Nagler[1], Jonas Zipfel[1], Christian Schüller[1], Tobias Korn[1], Alexey Chernikov[1], Rupert Huber[1†], Samuel Brem[2], Malte Selig[3], Gunnar Berghäuser[2] and Ermin Malic[2*]

[1] *Department of Physics, University of Regensburg, Regensburg, Germany*

[2] *Department of Physics, Chalmers University of Technology, Gothenburg, Sweden*

[3] *Department of Theoretical Physics, Technical University of Berlin, Berlin, Germany*

*Corresponding authors: †rupert.huber@ur.de, *ermin.malic@chalmers.se*

**Abstract**: Heterostructures of van der Waals bonded layered materials offer unique means to tailor dielectric screening with atomic-layer precision, opening a fertile field of fundamental research. The optical analyses used so far have relied on interband 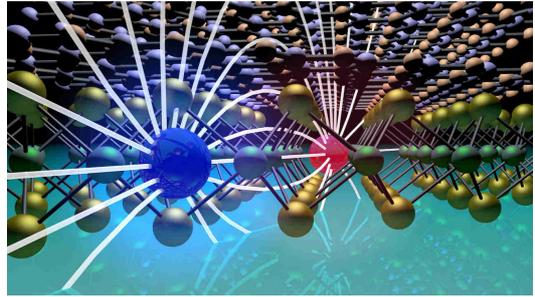 spectroscopy. Here we demonstrate how a capping layer of hexagonal boron nitride (hBN) renormalizes the internal structure of excitons in a $WSe_2$ monolayer using intraband transitions. Ultrabroadband terahertz probes sensitively map out the full complex-valued mid-infrared conductivity of the heterostructure after optical injection of $1s$ A excitons. This approach allows us to trace the energies and linewidths of the atom-like $1s$-$2p$ transition of optically bright and dark excitons as well as the densities of these quasiparticles. The fundamental excitonic resonance red shifts and narrows in the $WSe_2$/hBN heterostructure compared to the bare monolayer. Furthermore, the ultrafast temporal evolution of the mid-infrared response function evidences the formation of optically dark excitons from an initial bright population. Our results provide key insight into the effect of non-local screening on electron-hole correlations and open new possibilities of dielectric engineering of van der Waals heterostructures.

**Keywords**: dichalcogenides, atomically thin 2D crystals, van der Waals heterostructures, dielectric engineering, dark excitons.

**Main text**:

Atomically thin layers of transition metal dichalcogenides (TMDCs) have been in the spotlight of physical and chemical research due to a unique combination of properties. In particular, TMDC monolayers exhibit a direct energy gap in the optical range[1,2] – in sharp contrast to gapless graphene – which makes them fascinating candidates for ultimately thin optoelectronic devices[3-5]. Interestingly, the electronic and optical properties of this material system are dominated by Coulomb-bound electron-hole pairs, called excitons[6,7], rather than by unbound charge carriers. Reduced Coulomb screening in combination with the reduced dimensionality of atomically thin crystals leads to a dramatic increase of the exciton binding energy to typical values of few hundred meV, stabilizing these states even at room temperature[7-13]. Owing to the extreme confinement perpendicular to the plane of the material, excitons are particularly sensitive to the local environment surrounding the monolayer[14-22]. The possibility to vertically stack different cover materials onto two-dimensional TMDCs[4,23] has, thus, opened an exciting platform for fundamental physics[20,24-32] and device technologies[33,34]. On the one hand, heterostructures consisting of two different semiconducting monolayers have been employed to realize ultra-thin p–n junctions[33] and interlayer excitons[24,26,35] due to the possibility of interlayer charge transfer[24,26,33,35]. On the other hand, TMDC monolayers covered with an insulating van der Waals material have allowed for a less invasive strategy of Coulomb engineering[17,20]. By dielectric sculpting of the electric field lines connecting electrons and holes, in-plane electronic correlations can be modified within the TMDC monolayer without changing the chemical structure of the material itself[20]. Electron-phonon interactions[30,31,36,37] may be controlled and in-plane heterostructures can be realized on a nanometer length scale[17,20].

The most widespread methods of investigating excitons probe their interband generation or annihilation[7,10,11]. In these experiments, light couples dominantly to only a minor fraction of excitons, namely the optically bright ones. Optically dark states, whose interband dipole moment vanishes or whose center-of-mass momenta lie outside the light cone, cannot be directly addressed this way and have required more sophisticated in-plane geometries[38] or phonon-assisted processes[37]. Mid-infrared photons, in contrast, may directly induce a hydrogen-like 1$s$-2$p$ transition in pre-existing species, irrespective of interband dipole moments and large center-of-mass momenta. This concept has been employed to explore the formation time and dynamics of excitons in TMDC monolayers[12,13,39,40] and also represents a promising tool for the investigation of electronic correlations in heterostructures. Still, the direct influence of a cover layer on the correlations between orbital states of excitons has not been resolved in a direct, resonant way.

In this Letter, we experimentally investigate the influence of an insulating hexagonal boron nitride cover layer on the fundamental intra-excitonic 1$s$-2$p$ resonance by the absorption of mid-infrared photons in the TMDC monolayer $WSe_2$. By utilizing intraband spectroscopy, we are able to address all excitons in the system – including bright and dark states. This approach allows us to reveal the influence of the cover layer on the internal structure of excitons in the most direct possible way. We find that an hBN cover layer leads to a significant renormalization of the intra-excitonic 1$s$-2$p$ transition by 23 meV and a decrease of the transition linewidth compared to the uncovered monolayer. Moreover, the ultrafast evolution of the mid-infrared response functions shows a distinct blue shift of the intra-excitonic 1$s$-2$p$ resonance transition and an increase of its linewidth with time. Using microscopic modelling, we show that these observations are characteristic of the formation of dark excitons from the initial bright population.

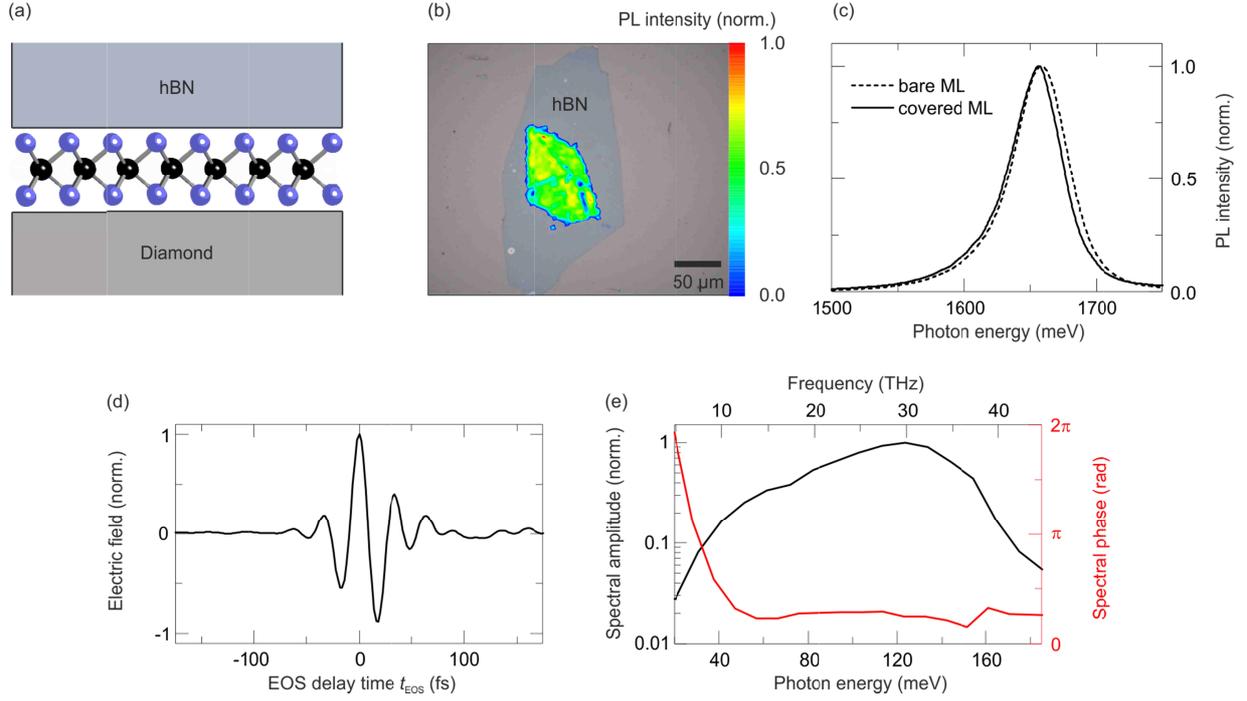

**Figure 1.** (a) Artist's view of the sample structure. A WSe$_2$ monolayer (blue and black spheres) on a CVD diamond window is covered with a capping layer of hexagonal boron nitride (hBN, thickness: 100 nm). The WSe$_2$/hBN heterostructure is fabricated by mechanical exfoliation of the individual components and subsequent deterministic transfer from a viscoelastic substrate onto a CVD diamond substrate. (b) Optical microscopy image and room-temperature photoluminescence intensity map of the WSe$_2$/hBN heterostructure excited by a continuous-wave laser at a wavelength of 532 nm. The lateral dimensions of the WSe$_2$ monolayer are about 100 μm x 75 μm. (c) Corresponding photoluminescence (PL) intensity spectrum (black solid curve) compared to the PL spectrum of a bare monolayer (black dashed curve). (d) Waveform of the phase-stable single-cycle mid-infrared probe pulse $E_{ref}$ transmitted through the unexcited heterostructure as a function of the electro-optic gate delay time $t_{EOS}$. (e) Corresponding multi-octave spanning amplitude (black curve) and phase spectrum (red curve) as obtained by Fourier transformation of $E_{ref}(t_{EOS})$.

The WSe$_2$/hBN heterostructure (Figure 1a) is manufactured by mechanical exfoliation and van der Waals bonding[41]. First, a monolayer of WSe$_2$ is deterministically transferred from a viscoelastic substrate onto a CVD diamond window. Subsequently, the cover layer is added onto the monolayer in the same way. The homogenous distribution of the photoluminescence (PL) intensity across the WSe$_2$ (Figure 1b) confirms the structural integrity of the monolayer after being covered with hexagonal boron nitride. Furthermore, the corresponding photoluminescence intensity spectrum (Figure 1c, black solid curve) features a peak energy of 1656 meV and a linewidth of 47 meV (FWHM). A comparison of these values to a PL spectrum of an uncovered monolayer (Figure 1c, black dashed curve) shows that covering reduces the linewidth of the PL by 5%, but hardly affects its spectral position. These observations are in line with recent interband spectroscopy results[18-20,22,35]. The thickness of the hBN cover layer is 100 nm, measured with atomic force microscopy.

To explore the influence of the capping layer on the electronic correlations in the WSe$_2$ monolayer, especially the fundamental intra-excitonic transition, we resonantly create bright 1s A excitons[12] by optical excitation with a 100 fs laser pulse centered at a wavelength of 745 nm (1664 meV, Supporting Information Figure S1). The pump fluence is set to $\Phi = 27$ µJ/cm$^2$ in order to keep the excitation density at a moderate level[40]. For the direct observation of the hydrogen-like 1s-2p transition of the injected excitons, we use phase-locked single-cycle mid-infrared (mid-IR) pulses (Figure 1d) generated by optical rectification of the fundamental laser output in a 10 µm-thin GaSe crystal. The pulses feature a flat spectral phase (Figure 1e, red curve) throughout their multi-octave spanning amplitude spectrum ranging from photon energies between 30 and 165 meV (Figure 1e, black curve). By electro-optic sampling up to the mid-infrared spectral region, the absolute amplitude and phase of the transmitted electric field waveform $E_{\text{ref}}(t_{\text{EOS}})$ are recorded as a function of the detection time $t_{\text{EOS}}$. We compare the mid-IR probe transient through the excited and the unexcited sample and determine the pump-induced change $\Delta E(t_{\text{EOS}})$ for a given delay time $t_{\text{PP}}$ between pumping and probing. A continuous variation of $t_{\text{PP}}$ provides access to the ultrafast dynamics of the non-equilibrium system (Supporting Information Figure S2). All experiments are performed at room temperature and ambient conditions.

From the quantities $E_{\text{ref}}(t_{\text{EOS}})$ and $\Delta E(t_{\text{EOS}}, t_{\text{PP}})$ the complex-valued mid-IR response function of the resonantly excited heterostructure can be extracted via a transfer matrix formalism[42,43]. The mid-IR response function includes the change in the real parts of the mid-IR conductivity ($\Delta\sigma_1$) and the dielectric function ($\Delta\varepsilon_1$), roughly corresponding to absorptive and inductive components, respectively. The advantage of the field-sensitive detection is that both functions can be extracted independently, without resorting to a Kramers-Kronig analysis.

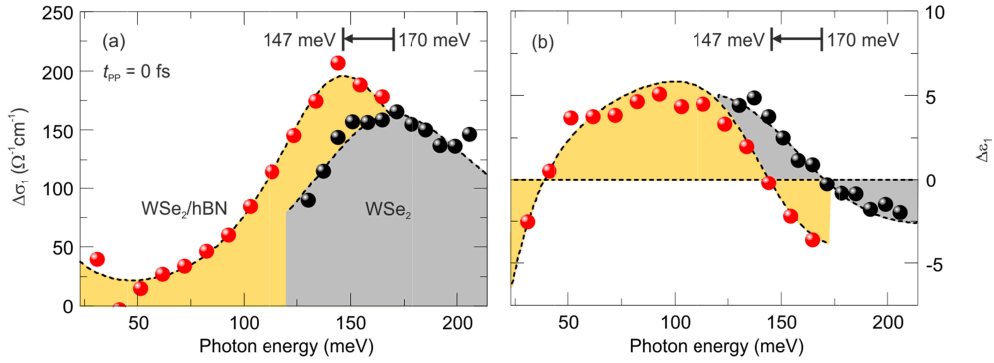

**Figure 2.** Pump-induced changes of the real parts of the optical conductivity $\Delta\sigma_1$ (a) and the dielectric function $\Delta\varepsilon_1$ (b) as a function of the photon energy, for a fixed pump delay time $t_{PP}$ = 0 fs after resonant excitation. The red spheres (yellow shading) denote the experimental data of the photoexcited WSe$_2$/hBN heterostructure (pump fluence $\Phi$ = 27 μJ/cm²), whereas the black spheres (gray shading) describe the experimental data of a photoexcited monolayer (pump fluence $\Phi$ = 25 μJ/cm², taken from ref 12). The black dashed curves represent the results by a phenomenological Drude-Lorentz model (eq 1) fitting simultaneously $\Delta\sigma_1$ and $\Delta\varepsilon_1$. This model takes unbound electron-hole pairs as well as excitons into account[13]. The arrows indicate the shift of the resonance from 170 meV (bare monolayer) to 147 meV (hBN covered monolayer).

Figure 2 compares the mid-IR response ($\Delta\sigma_1$ and $\Delta\varepsilon_1$) of the WSe$_2$/hBN *heterostructure* and a *bare uncovered* monolayer for a pump delay time of $t_{PP}$ = 0 fs. In case of the bare monolayer (Figure 2a,b, gray shading), a peak in $\Delta\sigma_1$ and a corresponding zero crossing in $\Delta\varepsilon_1$ around an energy of 170 meV are observed. These features are characteristic of a resonant absorption and can be assigned to the 1*s*-2*p* transition of excitons in this system[12,13]. In the case of the heterostructure (Figure 2a,b, yellow shading), we observe a similar spectral shape. However, the characteristic peak in $\Delta\sigma_1$ and the zero crossing in $\Delta\varepsilon_1$ are now located at a distinctly lower energy of 147 meV while the resonance is notably narrower. We will show below that these features originate from the renormalized 1s-2p transition of the resonantly generated excitons.

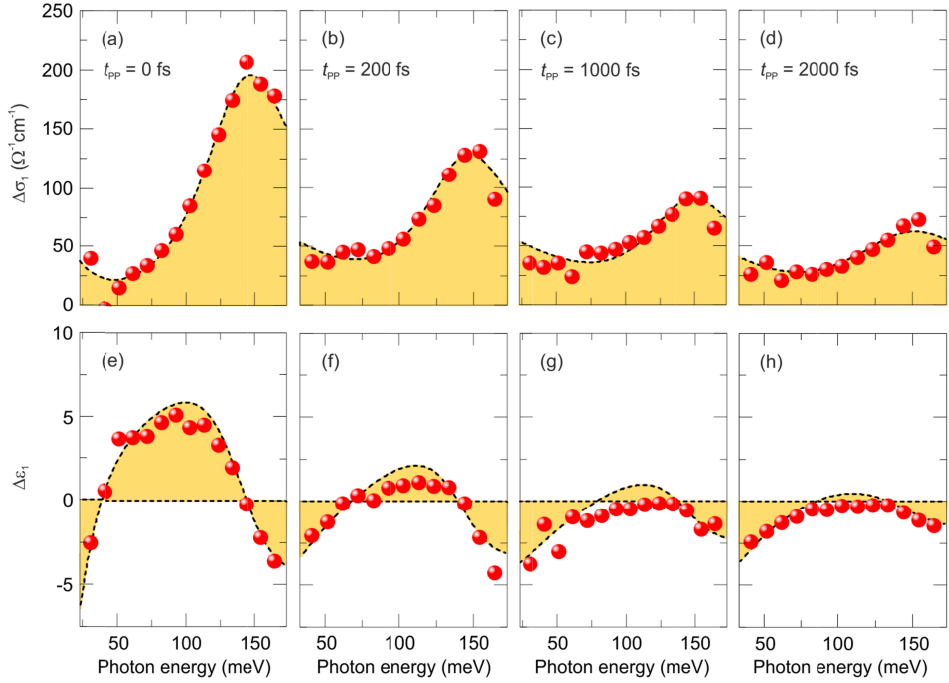

**Figure 3.** Pump-induced changes of the real parts of the optical conductivity $\Delta\sigma_1$ (a-d) and the dielectric function $\Delta\varepsilon_1$ (e-h) as a function of the photon energy, for several pump delay times $t_{PP}$ after resonant excitation. The red spheres (yellow shading) denote the experimental data of the photoexcited $WSe_2$/hBN heterostructure (pump fluence $\Phi = 27$ µJ/cm²). The black dashed curves represent the results of a phenomenological Drude-Lorentz model (eq 1) fitting simultaneously $\Delta\sigma_1$ and $\Delta\varepsilon_1$.

The nature of the screened excitons also manifests itself in their dynamics. Figure 3 displays the ultrafast evolution of the mid-IR response of the $WSe_2$/hBN heterostructure for a series of delay times $t_{PP}$. Compared to the response at $t_{PP} = 0$ fs (Figure 3a,e), the overall magnitude of the signal decreases, the resonance blue shifts slightly and the linewidth broadens as a function of $t_{PP}$ (Figure 3b-d,f-h). Interestingly, a second zero crossing in $\Delta\varepsilon_1$ persists on the low-frequency end of the spectral window, for all pump delay times $t_{PP}$. This feature may be caused by a second low-energy resonant absorption or by unbound electron-hole pairs, whose contribution would have a similar effect on $\Delta\varepsilon_1$.

For a more quantitative evaluation of the measured mid-IR response functions (Figures 2 and 3), we apply a Drude-Lorentz model[13,44,45] reflecting the contribution of excitons (Lorentz) and an electron-hole plasma (Drude) on a phenomenological level. Within this approach, pump-induced changes in the frequency-dependent dielectric function $\Delta\varepsilon(\omega) = \Delta\varepsilon_1 + i\Delta\sigma_1/(\varepsilon_0\omega)$ are described using two components:

$$\Delta\varepsilon(\omega) = \frac{n_X e^2}{d\varepsilon_0 \mu} \frac{f_{1s,2p}}{\frac{E_{res}^2}{\hbar^2} - \omega^2 - i\omega\Delta} - \frac{n_{FC} e^2}{d\varepsilon_0 \mu} \frac{1}{\omega^2 + i\omega\Gamma} \quad (1)$$

The first term, a Lorentzian resonance, represents the intra-excitonic 1$s$-2$p$ line. It includes the 1$s$ exciton density $n_X$, the corresponding reduced mass $\mu = m_e m_h / (m_e + m_h)$ ($m_e$ and $m_h$ are the effective masses of the constituent electron and hole), the effective thickness $d$ of the monolayer (treated as a thin slab in this model), the oscillator strength $f_{1s,2p}$ of the intra-excitonic transition, the resonance energy $E_{res}$, and the linewidth $\Delta$. Additional constants are the electron charge $e$ and the vacuum permittivity $\varepsilon_0$. The second term represents the Drude response of the electron-hole plasma, which depends on the density of free carriers $n_{FC}$ and their scattering rate $\Gamma$. For the analysis of the data, we fix the reduced mass $\mu = 0.17\, m_0$ and the oscillator strength for the covered monolayer $f_{1s,2p}^{hetero} = 0.30$ ($f_{1s,2p}^{bare} = 0.32$, see ref 12), corresponding to electron-hole properties at the K and K' valleys in WSe$_2$. The effective monolayer thickness $d$ is set to 0.7 nm. The remaining parameters of the Drude-Lorentz model ($n_X$, $n_{FC}$, $E_{res}$, $\Delta$, $\Gamma$) are extracted by fitting the experimental data. The fact that both independently measured $\Delta\sigma_1$ and $\Delta\varepsilon_1$ spectra need to be simultaneously reproduced poses strict limits on the possible values of the fitting parameters. The numerical adaptation (Figure 2 and 3, black dashed curves) yields an overall good fit quality. Note that this model has been well established for a phenomenological description of the intra-excitonic dielectric response[12,13,44,45], yet it neglects Coulomb correlations between unbound charge carriers as well as higher-order correlations between carriers and excitons[46]. Furthermore, it accounts only for a single excitonic resonance. In reality, however, multiple spectrally overlapping excitonic resonances can emerge for pump delay times $t_{PP} > 0$ fs (see discussion below). Equation 1 would model overlapping transitions by an increased broadening $\Delta$. Finally, excitonic resonances located far below the accessible spectral window yield a Drude-like contribution, which, in the framework of the above model, would be mimicked by the response of free carriers.

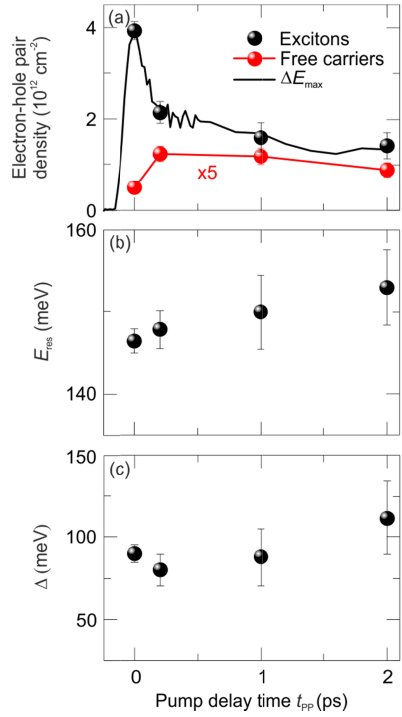

**Figure 4.** (a) Temporal evolution of the exciton density $n_X$ (black spheres, excitons) and the unbound electron-hole pair density $n_{FC}$ (red spheres, free carriers) as extracted from fitting the mid-IR response function of the WSe$_2$/hBN heterostructure (Figure 3, red spheres) measured for different pump delay times $t_{PP}$ with the Drude-Lorentz model. The pump-induced change $\Delta E$ recorded at fixed electro-optic sampling time $t_{EOS} = 0$ fs ($\Delta E_{max}$, black solid line) is proportional to the exciton density $n_X$. Resonance energy $E_{res}$ (b) and linewidth $\Delta$ (c) as a function of the pump delay time $t_{PP}$ as obtained from fitting the experimental mid-IR response data of the heterostructure (Figure 3, red spheres). The error bars represent the 95% confidence intervals of the fitting parameters.

For resonant excitation of the bare monolayer, the fitting procedure of the experimental data at a pump delay time $t_{PP}$ = 0 fs (Figure 2a,b, black spheres) yields an exciton density of $3.63 \times 10^{12}$ cm$^{-2}$, a resonance energy of 170 meV, and a linewidth of 117 meV. There is no measurable contribution from unbound electron-hole pairs in the investigated spectral region, in agreement with previous works[12,13]. In contrast, adding the hBN cover layer red shifts the resonance energy $E_{res}$ to 147 meV, whereas the linewidth $\Delta$ decreases to 90 meV at $t_{PP}$ = 0 fs (Figure 2a,b, red spheres). Furthermore, the fit of the heterostructure data reveals a non-vanishing density of unbound electron-hole pairs $n_{FC}$. Figure 4 summarizes the complete temporal evolution of the fitting parameters $n_X$, $n_{FC}$, $\Delta$ and $E_{res}$ for the WSe$_2$/hBN heterostructure. The density of excitons (Figure 4a, black spheres) initially decays within the first hundreds of femtoseconds from $n_X = 3.93 \times 10^{12}$ cm$^{-2}$ to $2.15 \times 10^{12}$ cm$^{-2}$ followed by a much slower decay on a picosecond scale, corresponding to the decay of coherent and incoherent exciton populations, respectively[47]. The pump-induced change $\Delta E$ observed at a fixed electro-optic sampling time $t_{EOS}$ = 0 fs (Figure 4a, black solid line) has been established to be proportional to the exciton density[12,13]. Thus, recording $\Delta E(t_{EOS} = 0$ fs$)$ as a function of $t_{PP}$ provides an alternative access to the temporal evolution of the exciton density. The almost perfect agreement of the temporal shape of $n_X(t_{PP})$ (Figure 4a, black spheres) with $\Delta E(t_{EOS} = 0$ fs, $t_{PP})$ (Figure 4a, black solid line) further corroborates the phenomenological Drude-Lorentz model of eq 1. The density of unbound electron-hole pairs $n_{FC}$ (Figure 4a, red spheres) rises within the first 200 fs, starting from a non-vanishing value $n_{FC} = 0.1 \times 10^{12}$ cm$^{-2}$, and slowly decays on a much longer timescale. Furthermore, $n_{FC}$ is an order of magnitude smaller than $n_X$ for all pump delay times $t_{PP}$. As noted above, the effect of a finite plasma density may also indicate the presence of a Lorentzian line located below the spectral window accessed here and will be further investigated in future research. The temporal evolution of the resonance energy $E_{res}$ is depicted in Figure 4b. Starting from an initial value of 147 meV at $t_{PP}$ = 0 fs, $E_{res}$ blue shifts up to 153 meV at $t_{PP}$ = 2 ps. The linewidth $\Delta$, however, decreases from 90 meV to 80 meV, within 200 fs, followed by a monotonic increase up to $\Delta$ = 112 meV at $t_{PP}$ = 2 ps (Figure 4c).

Both the spectral response and its ultrafast evolution provide novel evidence of renormalized Coulomb correlations shaped by the dielectric screening, as we show next. To explain the pronounced red shift of the intra-excitonic 1s-2p transition energy of the heterostructure by 23 meV, we perform calculations on a microscopic level explicitly taking into account the dielectric constant of the surrounding diamond substrate and the hBN cover layer. In the case of the heterostructure, we expect increased screening of the Coulomb potential compared to the bare monolayer, which should result in a reduction of the binding energy of excitons and, thus, in a red shift of the 1s-2p transition energy. Solving the excitonic Wannier equation in momentum representation[36]

$$\frac{\hbar q^2}{2\mu}\Psi_n(\boldsymbol{q}) - \sum_k V_{\text{exc}}(\boldsymbol{q},\boldsymbol{k})\,\Psi_n(\boldsymbol{q}+\boldsymbol{k}) = E_n\Psi_n(\boldsymbol{q}) \qquad (2)$$

we have microscopic access to exciton binding energies $E_n$ and wave functions $\Psi_n(\boldsymbol{q})$ with index $n = 1s$, $2s$, $2p$, … . Here, we have introduced the excitonic part of the Coulomb interaction $V_{\text{exc}}(\boldsymbol{q},\boldsymbol{k})$, which is treated within the thin-film formalism for 2D systems[8,9,14,48,49]. The latter also contains the dielectric background screening $\varepsilon_{\text{BG}} = \frac{\varepsilon_{\text{cover}}+\varepsilon_{\text{substrate}}}{2}$ of the underlying substrate/cover layer. Figure 5 shows the calculated binding energies and wave functions ($n = 1s$, $2p$) for bright ($Q = 0$) as well as momentum-forbidden dark excitons ($Q = \Lambda$) as a function of the center-of-mass momentum $Q$, comparing the situation of the bare and covered monolayer. We can perfectly reproduce the experimentally observed 1s-2p resonance energy of 147 meV with a dielectric constant of $\varepsilon_{\text{cover}} = 2.3$, which agrees well with the literature value for hBN[50]. Thus, we can unequivocally assign the observed resonance at 147 meV to the renormalized 1s-2p transition in the heterostructure.

The ultrafast evolution of the resonance energy $E_{\text{res}}$ shows a monotonic blue shift with increasing pump delay time $t_{\text{PP}}$ (Figure 4b). This feature can be explained by the formation of a momentum-forbidden dark exciton population via exciton-phonon scattering or the emergence of spin-unlike excitons due to spin-flip processes[40]. Since the intra-excitonic 1s-2p separation of dark excitons is higher than the one of bright excitons (Figure 5a), $E_{\text{res}}$ increases as the dark population forms[40]. Furthermore, the dark excitonic ground state ($Q = \Lambda$) in WSe$_2$ is energetically lower than the bright one (Figure 5a) leading to an efficient accumulation of excitons in the $\Lambda$-valley during thermalization[40].

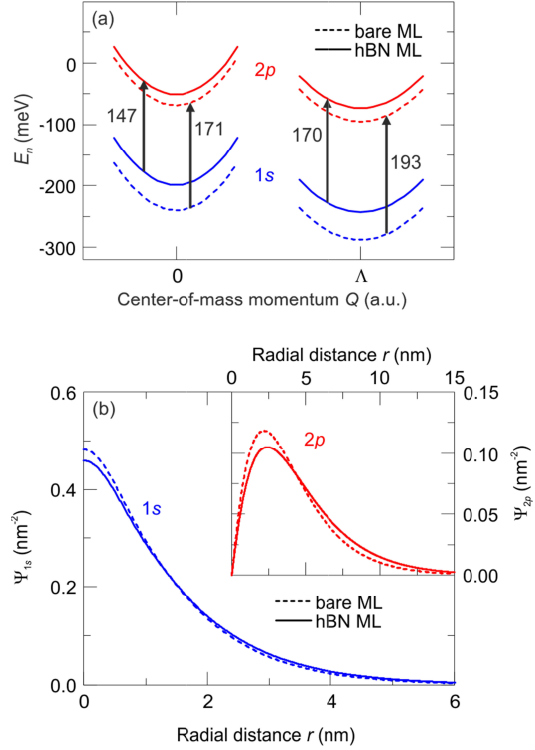

**Figure 5.** (a) Calculated excitonic binding energies $E_n$ ($n = 1s$, $2p$) for bright ($Q = 0$) as well as momentum-forbidden dark excitons ($Q = \Lambda$) as a function of the center-of-mass momentum $Q$. The black arrows symbolize various intra-excitonic $1s$-$2p$ transitions. (b) Calculated radial wave functions $\Psi_n(r)$ for the $1s$ (blue curves) and $2p$ (red curves) bright excitonic levels. The dashed lines are related to the bare monolayer, whereas the solid lines refer to calculations for the monolayer covered with hBN.

Next we turn to the strong reduction of the linewidth $\Delta$ from 117 meV for the bare monolayer to 90 meV for the heterostructure at $t_{PP} = 0$ fs. This fact appears even more striking when the corresponding exciton densities are considered. $\Delta$ is expected to increase with $n_X$[12,13]. In our experiment, the linewidth $\Delta_{hetero}$ is found to be smaller than $\Delta_{bare}$ by 23%, even though the density $n_X^{hetero}$ slightly exceeds $n_X^{bare}$ by 8%. Thus, the hBN cover layer clearly reduces the width of the intra-excitonic 1s-2p transition line. One of the possible broadening mechanisms contributing to the linewidth $\Delta$ is exciton-phonon scattering. Here, the scattering efficiency is i.a. determined by the overlap of excitonic wave functions[36]. Since stronger Coulomb screening, caused by the change of the dielectric environment, leads to more spatially extended exciton wave functions (Figure 5b), exciton-phonon scattering should be modified in the heterostructure. However, our numerical calculations reveal that in the relevant range of dielectric background constants $\varepsilon_{BG}$, the scattering with acoustic phonons is slightly enhanced by adding hBN, whereas the interaction with optical phonons becomes weaker, so that in total the exciton-phonon scattering is almost the same with and without hBN. Thus, this mechanism cannot cause the observed decrease of the linewidth $\Delta$. Exciton-exciton scattering, however, depends on the Coulomb potential which is indirectly proportional to $\varepsilon_{BG}$. Thus, the increase of $\varepsilon_{BG}$, caused by adding a cover layer, can lead to a decrease of exciton-exciton scattering, resulting in a smaller linewidth $\Delta$.

The temporal evolution of the linewidth (Figure 4c) shows another unexpected behavior. The linewidth $\Delta$ should decrease as a function of $t_{PP}$ since the overall electron-hole pair density decreases (Figure 4a as well as Supporting Information Figures S3 and S4) and thus fewer scattering events take place. However, the linewidth – after an initial and expected decrease – starts to rise from 80 meV at $t_{PP} = 0.2$ ps to 112 meV at $t_{PP} = 2$ ps. This anomalous behavior is a second indication for dark excitons that are not located at the K/K'-points of the Brillouin zone and thus have different resonance energies[40]. Since our mid-infrared probe pulse is sensitive to all kinds of dark excitons, the superposition of various intra-excitonic 1s-2p absorption peaks from different momentum- or spin-forbidden states leads to an apparent increase of the linewidth, as the dark population ($t_{PP} > 0.2$ ps) forms from the initial bright exciton population ($t_{PP} < 0.2$ ps). Since the overall electron-hole pair density steadily decreases for $t_{PP} > 0.2$ ps (Figure 4a) and thus the linewidth $\Delta$ should show a similar trend, the observed strong blue shift of $E_{res}$ can only be accounted for by the formation of dark excitons[40].

In conclusion, intraband spectroscopy in the mid-infrared spectral range allows us to investigate the influence of the modification of the dielectric environment on bright and dark excitons in a WSe$_2$ monolayer. The data quantify internal transition energies, densities and many-body effects. Most remarkably, we observe a significant red shift of the intra-excitonic 1*s*-2*p* transition as well as a decrease of its linewidth compared to the uncovered monolayer. Furthermore, the distinct blue shift of the 1*s*-2*p* resonance along with the anomalous dynamics of the linewidth suggests the formation of a dark exciton population from an initial bright one. These heterostructures fabricated by simple stacking offer an exciting playground to custom-tailor electronic correlations and thus open exciting perspectives for novel optoelectronic applications. In future studies one may even start to resonantly hybridize excitonic transitions with low-energy excitation of surrounding cover layers, designing novel material parameters by layer-sensitive dielectric sculpting.

**Acknowledgements:**

The authors thank Matthias Knorr for helpful discussions, Kenji Watanabe and Takashi Taniguchi for providing high quality hBN flakes, as well as Martin Furthmeier for technical assistance. This work was supported by the European Research Council through ERC grant 305003 (QUANTUMsubCYCLE) and by the Deutsche Forschungsgemeinschaft (DFG) through Research Training Group GRK 1570 and project grant KO3612/1-1. AC gratefully acknowledges funding from the Deutsche Forschungsgemeinschaft through the Emmy Noether Programme (CH1672/1-1). The Chalmers group acknowledges funding from the European Union's Horizon 2020 research and innovation program under grant agreement No. 696656 (Graphene Flagship) and the Swedish Research Council (VR). The Berlin group was supported by the Deutsche Forschungsgemeinschaft (DFG) through the collaborative research centers SFB 787 and 951.

**Supporting Information:**

1. Interband absorbance of the $WSe_2$/hBN heterostructure
2. Field-resolved optical pump/mid-infrared probe spectroscopy
3. Pump fluence dependence of the linewidth of the intra-excitonic $1s$-$2p$ transition